\documentclass[a4paper,12pt]{article}

\usepackage{amsmath, amssymb, a4wide, bm, cancel, color, graphicx, subfig, enumerate, appendix, cite, hyperref}

\newcommand{\Tr}[2]{\mathop{\rm Tr}_{#1}\left[#2\right]}
\newcommand{\diff}[2]{\frac{{\rm d}#1}{{\rm d}#2}}
\newcommand{\pdiff}[2]{\frac{\partial#1}{\partial#2}}
\newcommand{\dimint}[2]{\int\mathrm{d}^{#1}#2\,}
\newcommand{\dimintlim}[4]{\int_{#3}^{#4}\mathrm{d}^{#1}#2\,}
\newcommand{\nbrack}[1]{\left(#1\right)}
\newcommand{\sbrack}[1]{\left[#1\right]}
\newcommand{\expect}[1]{\langle#1\rangle}

\newcommand{\sep}[1]{\quad\mbox{#1} \quad}

\def\widerow{\rule{0pt}{2.5ex}\rule[-1.5ex]{0pt}{0pt}}
\def\be{\begin{equation}}
\def\ee{\end{equation}}
\def\ba{\begin{eqnarray}}
\def\ea{\end{eqnarray}}
\def\cF{{\cal F}}
\def\cL{{\cal L}}
\def\cM{{\cal M}}
\def\cN{{\cal N}}
\def\cO{{\cal O}}
\def\uno{\mbox{1 \kern-.59em {\rm l}}}

\numberwithin{equation}{section}
\numberwithin{figure}{section}
\numberwithin{table}{section}

\begin{document}

\title{{\normalsize DCPT-11/19\hfill\mbox{}\hfill\mbox{}}\\
\vspace{2.5 cm}
\Large{\textbf{Condensate cosmology in O'Raifeartaigh models}}}
\vspace{2.5 cm}
\author{James Barnard\\[3ex]
\small{\em Department of Mathematical Sciences,}\\
\small{\em Durham University, Durham DH1 3LE, UK}\\[1.5ex] 
\small james.barnard@durham.ac.uk\\[1.5ex]}
\date{}
\maketitle
\vspace{2ex}
\begin{abstract}
\noindent Flat directions charged under an R-symmetry are a generic feature of O'Raifeartaigh models.  Non-topological solitons associated with this symmetry, R-balls, are likely to form through the fragmentation of a condensate, itself created by soft terms induced during inflation.  In gravity mediated SUSY breaking R-balls decay to gravitinos, reheating the universe.  For gauge mediation R-balls can provide a good dark matter candidate.  Alternatively they can decay, either reheating or cooling the universe.  Conserved R-symmetry permits decay to gravitinos or gauginos, whereas spontaneously broken R-symmetry results in decay to visible sector gauge bosons.
\end{abstract}

\section{Introduction and conclusions}

Of upmost importance in any supersymmetric theory is the appearance of classical flat directions, or pseudo-moduli: scalar combinations along which the classical potential is invariant.  Indeed, ref.~\cite{Intriligator:2008fe} demonstrates that the quantum mechanical stabilisation (or not) of flat directions is instrumental in determining whether a SUSY breaking vacuum is viable.  Furthermore any renormalisable O'Raifeartaigh model \cite{ORaifeartaigh:1975pr} necessarily has such a direction, namely the scalar component of the goldstino superfield \cite{Ray:2006wk, Komargodski:2009jf}.

Of course, flat directions are well known to exist in the MSSM \cite{Enqvist:2003gh} and are frequently utilised when explaining the source of the universe's baryon asymmetry \cite{Affleck:1984fy}.  High scale SUSY breaking associated with inflation can result in a condensate with large expectation value forming along a flat direction charged under baryon number.  If the model does not preserve it at high energies, this condensate carries significant fractional baryon number that survives to the present day.  Originally it was thought that the condensate would evaporate into light fermions (imparting its baryon number to visible sector degrees of freedom) but later studies \cite{Turner:1983he, Kusenko:1997si, Enqvist:1997si, Kasuya:1999wu, Kasuya:2000wx, Kasuya:2001hg, Multamaki:2002hv, Jokinen:2002xw} instead suggest that the condensate fragments into localised lumps known as Q-balls \cite{Coleman:1985ki}.

Given that O'Raifeartaigh models have flat directions of their own, one can ask whether something similar happens in the hidden sector.  Recent work \cite{Shih:2009he, KerenZur:2009cv, Amariti:2009tu, Bell:2011tn, Cheung:2011if} has certainly suggested that other, non-MSSM flat directions can have important consequences.  Nelson and Seiberg's powerful argument \cite{Nelson:1993nf} implies that a generic, calculable model can only break $\cN=1$ SUSY if it has a (possibly spontaneously broken) U(1) R-symmetry.  Even if the SUSY breaking is only metastable one expects an approximate R-symmetry, with any R-violating operators being suppressed to ensure a sufficiently long lived vacuum.  Rather than baryon number, it is this R-charge that is carried by flat directions in O'Raifeartaigh models, thus leading to {\em R-balls}: non-topological solitons stabilised by a global U(1) R-symmetry\footnote{This may not be the only bearing global symmetries have on the cosmology of the SUSY breaking sector.  Ref.~\cite{Barnard:2010wk} entertains the possibility of soliton induced, SUSY restoring phase transitions in models of metastable SUSY breaking.}.

In this paper we investigate the general, cosmological evolution of flat directions in O'Raifeartaigh models.  Since many such models of SUSY breaking are only low energy effective descriptions, it is important to consider effects from the underlying microscopic theory.  The cutoff scale of the low energy theory will prove to be an important parameter.  Initially, the flat directions evolve similarly to their MSSM counterparts and a spatially constant condensate is formed.  As the universe cools down the condensate fragments (subject to constraints imposed by thermal effects) and forms R-balls.  At this stage it doesn't make a difference whether R-symmetry is spontaneously broken or not.  The effective potential for the flat direction preserves an R-symmetry over the majority of its domain and only small field values, i.e.\ the edges of the R-ball, see the effects of any breaking.  Eventually R-balls decay, either by surface evaporation to light fermions if R-symmetry is preserved, or to light bosons throughout the interior if R-symmetry is spontaneously broken.

Two distinct phenomena can then be attributed to R-balls.  If sufficiently long lived they act as cold dark matter and are easily capable of providing the dominant contribution to the observed density.  This scenario only occurs in gauge mediated SUSY breaking where the SUSY breaking scale (or equivalently the gravitino mass) is small enough for R-balls to outlast the current age of the universe.  Otherwise R-balls decay.  A significant density during decay leads to a second bout of reheating or a cooling of the universe, depending on the temperature of the decay products.  Allowing the universe to pass through an R-ball dominated epoch thus decouples the dynamics of the inflaton from the generation of visible sector matter.  Gravity mediated SUSY breaking sees R-balls evaporate to gravitinos, reheating the universe.  Gauge mediation allows for both reheating and cooling, by gravitinos and gauginos if R-symmetric, and by visible sector gauge bosons when R-breaking.  The parameter space of some example O'Raifeartaigh models is dissected in figures \ref{fig:RPgr}, \ref{fig:RPgaRS}, \ref{fig:RPgaRB}.  In each case there remain several regions where R-balls are forbidden, either by thermal effects or due to contradiction with experimental evidence, or form but have little obvious effect.

The remainder of the paper is organised as follows.  In section \ref{sec:FD} we will review the observation of refs.~\cite{Ray:2006wk, Komargodski:2009jf} regarding the existence of flat directions in O'Raifeartaigh models, and will discuss how they are lifted by loop corrections.  Section \ref{sec:CE} deals with the cosmological evolution of the flat direction, from inflation to fragmentation, including potential thermal effects.  R-balls themselves are investigated in section \ref{sec:RB}, with a particular emphasis on their decay rates.  Finally, in section \ref{sec:RP}, we will consider the phenomenological consequences.  Throughout this work an attempt is made to keep the analysis as general as possible.  It is inevitable that certain aspects (decay, for example) would benefit from a more specific, numerical treatment.  However, the aim here is to broadly consider the entire spectrum of consequences associated with flat directions in O'Raifeartaigh models.

\section{Flat directions in O'Raifeartaigh models\label{sec:FD}}

To begin we will recap the results of refs.~\cite{Ray:2006wk, Komargodski:2009jf} on flat directions in O'Raifeartaigh models, followed by a discussion on their lifting.  Consider a model of SUSY breaking which is generic, calculable and has a low energy description in the form of an O'Raifeartaigh model.  Ref.~\cite{Nelson:1993nf} tells us that this model must have an R-symmetry.  Denoting the chiral superfields as $\{\Phi_a\}$ the scalar potential is given by the F-terms
\be
U(\Phi)=F_a^\dag F_a=W_a^\dag W_a
\ee
where $W_a$ is the derivative $\partial W/\partial\Phi_a$.  For a locally stable SUSY breaking vacuum to exist we must be able to find a solution $\Phi_a=\expect{\Phi}_a$ to the equations
\begin{align}\label{eq:FDFterms}
\pdiff{U}{\Phi_a}=W_{ab}W_b^\dag & =0 & W_a\neq0\quad\mbox{for at least one value of }a.
\end{align}
The first equation contains the fermion mass matrix $\cM_{1/2}\equiv W_{ab}$ so these two statements confirm that there is at least one massless fermion, the goldstino, corresponding to the fermionic component of the chiral superfield whose F-term gets a VEV.  Meanwhile the scalar mass matrix is
\be\label{eq:IBcM0}
\cM_0^2=\nbrack{\begin{array}{cc}
\cM_{1/2}^\dag \cM_{1/2} & \cF^\dag \\
\cF & \cM_{1/2}\cM_{1/2}^\dag \end{array}}\sep{where}
\cF_{ab}=W_{abc}W_c^\dag
\ee
and must be positive semi-definite for $\Phi_a=\expect{\Phi}_a$ so that the vacuum is free of tachyons.

Now take the direction $X_a=W_a^\dag|_{\Phi_a=\expect{\Phi}_a}$ (the scalar component of the goldstino superfield) and consider the norm
\be
\nbrack{\begin{array}{cc}X^\dag & X^T\end{array}}\cM_0^2\nbrack{\begin{array}{c}X \\ X^*\end{array}}=
X^T\mathcal{F}X+\mbox{h.c.}
\ee
This quantity necessarily vanishes.  Otherwise one could always rotate the phase of $X$ to make the right hand side negative, contradicting the assumption that the vacuum is locally stable.  Ergo it must be that $\mathcal{F}X=0$, therefore $(X,X^*)$ is a massless scalar.  In fact the claim is much stronger than this for a renormalisable superpotential.  $\mathcal{F}X=0$ can be expanded as
\be\label{eq:FDFX0}
(\mathcal{F}X)_a=\sbrack{W_{abc}W_b^\dag W_c^\dag}_{\Phi_a=\expect{\Phi}_a}=0.
\ee
Moving along the $X$ direction the scalar potential changes as
\be
U(\expect{\Phi}+\Delta X)=U(\expect{\Phi})+\delta W_a^\dag\delta W_a
\ee
for some complex parameter $\Delta$, where
\be
\delta W_a=\sbrack{\Delta W_{ab}W_b^\dag+\frac{1}{2}\Delta ^2W_{abc}W_b^\dag W_c^\dag}_{\Phi_a=\expect{\Phi}_a}.
\ee
This expansion is exact up to fourth derivatives in $W$ (i.e.\ for a renormalisable superpotential) and, upon consulting eqs.~\eqref{eq:FDFterms} and \eqref{eq:FDFX0}, one immediately finds $\delta W_a=0$.  In other words the scalar potential is invariant under complex translations along the scalar component of the goldstino superfield: it is a flat direction.  For non-renormalisable models the scalar partner of the goldstino remains massless, but the degeneracy may be lifted by higher order terms in the superpotential.

A consequence is that any O'Raifeartaigh model can be recast in terms of a goldstino superfield $X$, whose F-term VEV is responsible for SUSY breaking, and some other superfields $\{\varphi_i\}$ \cite{Komargodski:2009jf}.  In this basis we will allow the most general, renormalisable form for the superpotential
\be\label{eq:FDW}
W=fX+(\mu_{ij}+\lambda_{ij}X)\varphi_i\varphi_j+\kappa_{ijk}\varphi_i\varphi_j\varphi_k
\ee
for coupling constants $f$, $\mu$, $\lambda$ and $\kappa$, where $f$ is assumed real and positive without loss of generality.  Note that these symbols will frequently be used without their indices to denote the generic size of the couplings.  The superfields $\varphi_i$ are defined so as to have vanishing VEVs whereas the scalar component of $X$ is our classical flat direction.  Transforming into this basis may not respect the global symmetry group of the model but, in a model with an R-symmetric vacuum, $X$ clearly has R-charge $+2$.  Any renormalisable O'Raifeartaigh model with an R-symmetric vacuum therefore possesses a flat direction with non-zero R-charge.

\subsection{Lifting the flat direction}

For the vacuum to be well defined $X$ cannot remain flat and must be stabilised.  At low energy this is accomplished by quantum effects, typically at one loop via the Coleman-Weinberg potential \cite{Coleman:1973jx} evaluated with respect to a UV cutoff scale $M$.  Around this scale we may start to see non-renormalisable effects from an underlying microscopic theory, which could also lift the flat direction, but are highly suppressed at low energy so have little impact.  Exactly where the Coleman-Weinberg potential is minimised depends on the details of the model but one can deduce its approximate form in two limits.  Close to its minimum at $X=\expect{X}$ the effective potential goes like
\be\label{eq:FDUeffs}
U_{\rm eff}(X)=U_0+\frac{1}{2}m^2|X-\expect{X}|^2+\cO(|X-\expect{X}|^3).
\ee
If R-symmetry is preserved this minimum must be at $\expect{X}=0$.  The mass term $m^2$ depends on the couplings $f$, $\mu$ and $\lambda$ and one can deduce its value either by explicitly calculating the Coleman-Weinberg potential for a given model, or by using the more elegant methods developed in ref.~\cite{Shih:2007av}\footnote{It was also shown in ref.~\cite{Shih:2007av} that spontaneous R-symmetry breaking requires a superfield with R-charge not equal to $0$ or $+2$.}.  Schematically it should go like
\be\label{eq:FDm}
m^2\sim\frac{\lambda^4f^2}{16\pi^2\mu^2}.
\ee

At the other end of the scale when $X$ is large, specifically $|\mu+\lambda X|^2\gg\lambda f$, the SUSY breaking is small.  In fact an appeal to naturalness suggests $f$ and $\mu^2$ should not be too dissimilar so small SUSY breaking often implies $|\lambda X|\gg|\mu|$.  In this limit we can find an alternative formulation for the effective potential by integrating out the massive $\varphi$'s.  The process contributes an extra term to the K\"ahler potential \cite{Intriligator:2006dd}
\be
K_{\rm eff}(X,X^\dag)=|X|^2-\frac{1}{32\pi^2}\Tr{}{\cM^\dag\cM\ln{\nbrack{\frac{\cM^\dag\cM}{M^2}}}}
\ee
and consequently leads to a one loop effective potential
\be\label{eq:FDUeffl}
U_{\rm eff}(X)=\frac{|F_X|^2}{\partial_X\partial_{X^\dag}K_{\rm eff}}
=f^2\nbrack{1+\frac{1}{16\pi^2}\sum_i|\lambda_i|^2\sbrack{1+\ln{\nbrack{\frac{|\lambda_iX|}{M}}}}+\cO(\lambda^4)}.
\ee
The mass matrix $\cM=\lambda X$ has been taken from the superpotential \eqref{eq:FDW} (neglecting $\mu$ in the limit of small SUSY breaking) where $\{\lambda_i\}$ are the eigenvalues of $\lambda$.  If $\mu^2\gg f$ for some reason, the K\"ahler potential is modified by the possible appearance of additional light states at certain points in the pseudo-moduli space, which cannot be integrated out.  These lead to singular behaviour and our approximation must be reassessed.  Regardless, all conclusions reached apply wherever $|\mu|\ll|\lambda X|\ll M$.  Note also that eq.~\eqref{eq:FDUeffl} is only valid up to second order in $f$ and $\lambda$.  If the $\lambda$'s are large, or there are a vast number of non-goldstino superfields, higher loop corrections become important and invalidate the Coleman-Weinberg formula.

\begin{figure}[!tb]
\begin{center}
\includegraphics[width=0.44\textwidth]{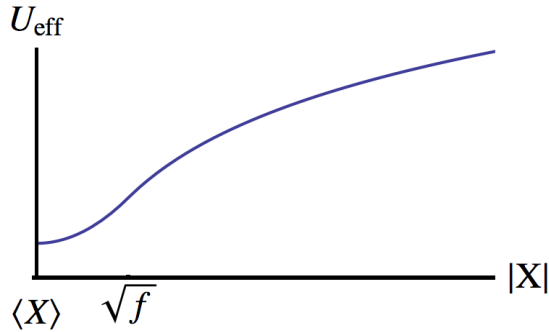}
\caption{A typical low energy effective potential for the scalar component of the goldstino superfield $X$.  At tree level $X$ is a flat direction but is lifted by loop effects.  For $X^2\ll f$ these are polynomial in nature and stabilise $X$ at some value $\expect{X}$ (equal to zero in an R-symmetric vacuum).  For $X^2\gg f$ the quantum corrections become logarithmic.\label{fig:FDUeff}}
\end{center}
\end{figure}

In the following we will be interested only in the potential relative to the SUSY breaking vacuum so will omit the constant term of eq.~\eqref{eq:FDUeffs}, which goes like $f^2$ up to loop effects.  Normalising both potentials thus gives the final form
\begin{align}\label{eq:FDUeff}
U_{\rm eff}(X) & \approx\frac{1}{2}m^2|X-\expect{X}|^2 && \mbox{for } X^2\ll f \nonumber\\
U_{\rm eff}(X) & \approx\frac{f^2}{16\pi^2}\sum_i|\lambda_i|^2\sbrack{\zeta +\ln{\nbrack{\frac{|\lambda_iX|}{M}}}} && \mbox{for } f\ll X^2\ll M^2
\end{align}
for some order one parameter $\zeta$.  For a model that is perturbatively well behaved and whose couplings satisfy $f\sim\mu^2$ this effective potential is illustrated in figure \ref{fig:FDUeff}.

\section{Cosmological evolution\label{sec:CE}}

Flat directions in the MSSM have been extensively studied (see e.g.\ ref.~\cite{Enqvist:2003gh} and the references therein) and many of the techniques used are directly applicable here.  In particular, a condensate forms along flat directions in a wide variety of models due to the Affleck-Dine mechanism \cite{Affleck:1984fy}.  The idea is that quantum fluctuations are spread out during an inflationary period of the universe's evolution, with only the long wavelength modes surviving to form a spatially constant condensate.  Initially, the expectation value of the condensate field is set by high scale, SUSY breaking couplings to the field driving inflation: the inflaton.  As the universe cools down, the expectation value of the condensate decreases until inflaton effects become subdominant and the condensate field begins to move in the low energy effective potential.  At this point the condensate becomes unstable to spatial perturbations and begins to fragment \cite{Turner:1983he, Kusenko:1997si, Enqvist:1997si, Kasuya:1999wu, Kasuya:2000wx, Kasuya:2001hg, Multamaki:2002hv, Jokinen:2002xw}, the final state being non-topological solitons stabilised by some conserved charge: Q-balls \cite{Coleman:1985ki}.

In O'Raifeartaigh models the overall picture is similar.  We will keep with standard nomenclature and refer to the flat direction $X$ as the {\em condensate field}.  A summary of the key points is then as follows.
\begin{itemize}
\item{\bf Inflation:}  A tachyonic soft mass, originating from the high scale SUSY breaking driving inflation, can drive the condensate field away from its low energy VEV.  The condensate field is stabilised by higher order soft terms originating from a microscopic theory.   For an accidental R-symmetry the condensate is stabilised at a scale $\Lambda$, parametrically between $M$ and $\sqrt{f}$, whereas for an exact R-symmetry it is stabilised around the Planck scale.
\item{\bf Rotation:}  The Hubble parameter decreases until the inflaton induced soft terms are comparable in scale to the low energy effective potential.  The low energy vacuum is restored and the condensate field begins to rotate in its potential well with magnitude $\Lambda$ and frequency $\lambda f/4\pi\Lambda$.
\item{\bf Fragmentation:}  Shortly after beginning rotation the condensate is rendered unstable to spatial perturbations and fragments into localised lumps of size $2\pi\sqrt{2}\Lambda/\lambda f$.  The fragments coalesce into extended, classical objects with large charge: R-balls.
\item{\bf Decay:} In R-symmetric models R-balls evaporate to gravitinos at tree level and other light fermions at one loop.  In models with spontaneously broken R-symmetry R-balls can decay more quickly into other light fermions at tree level and also into light bosons.
\end{itemize}
We shall now discuss each stage of the evolution in more detail.

\subsection{Inflation}

At such early epochs one cannot consider the low energy O'Raifeartaigh model in isolation.  Inflation induces extra soft terms in the potential related to the Hubble parameter \cite{Dine:1995kz}.  Non-renormalisable operators arising from the microscopic theory also have a profound effect on the early universe dynamics, especially if they do not respect the R-symmetry of the low energy theory.

Cosmologically the scalar component of $X$ obeys the equation of motion
\be\label{eq:CEeom}
\ddot{X}+3H\dot{X}-\frac{1}{a^2}\nabla^2X+\pdiff{V_{\rm eff}}{X^\dag}=0
\ee
for a high energy effective potential $V_{\rm eff}(X)$.  $H$ is the Hubble parameter, $a$ is the scale factor of the universe and, for a homogeneous condensate, the gradient term obviously disappears.  In the absence of thermal effects (the discussion of which is postponed until section \ref{sec:CEtherm}) the general form of $V_{\rm eff}(X)$ is known \cite{Dine:1995kz} to be
\be\label{eq:CEVeff}
V_{\rm eff}(X)=-cH^2|X|^2+\frac{H}{M^{n-3}}\nbrack{A\eta X^n+\mbox{h.c.}}+\frac{1}{M^{2n-6}}|\eta|^2|X|^{2n-2}
\ee
for a microscopic superpotential coupling $\eta$, order one constants\footnote{For D-term inflation one expects $A=0$.  This has little effect for an accidental R-symmetry but, when R-symmetry is exact, D-term inflation precludes the possibility of R-violating operators and the condensate always has vanishing charge.} $c$ and $A$, and where $n\ge4$.  The first term is a soft mass induced through inflaton couplings and is always present.  The remaining terms are generated by non-renormalisable superpotential operators originating from the microscopic theory, which has been allowed to break R-symmetry (i.e.\ the R-symmetry of the low energy theory is accidental) and lift the flat direction.  Even if R-symmetry is exact up to the Planck scale it is expected to be broken by gravitational effects.  If no non-renormalisable superpotential operators lift the flat direction, one would instead have
\be
V_{\rm eff}(X)=-cH^2|X|^2+\frac{H^2}{M_P^{n-3}}AX^{n-1}+\frac{H^2}{M_P^{2n-2}}B|X|^{2n-4}
\ee
but the conclusions of the subsequent discussion are unaffected, other than replacing the scale $\Lambda$ (defined shortly) with the Planck scale $M_P$.

The soft mass in eq.~\eqref{eq:CEVeff} is extremely important in the early universe when $H$ is large.  For the minimal K\"ahler potential $K=X^\dag X$ it is generated by supergravity corrections of the form
\be
V_{\rm eff}(X)=e^{K/M_P^2}V_{\rm inf}(\chi)
\ee
where $\chi$ is the inflaton superfield.  During inflation the inflaton vacuum energy dominates the universe so $V_{\rm inf}(\chi)\sim H^2M_P^2$, leading to a soft mass term with negative $c$.  Therefore the effective potential is minimised at the origin and the dynamics are uninteresting.  However, for a non-minimal K\"ahler potential it is quite possible that $c$ is positive.  Consider, for example, the term $K\supset(\chi^\dag\chi)(X^\dag X)/M_P^2$ which is allowed by all possible symmetries of the model.  In fact terms of this form are inevitable for superpotentials like \eqref{eq:FDW}: they arise as counterterms for Yukawa couplings \cite{Gaillard:1993es, Bagger:1995ay, Dine:1995kz}.  Their contribution to the soft mass is
\be
\delta\cL=\dimint{4}{\theta}\frac{(\chi^\dag\chi)(X^\dag X)}{M_P^2}=\frac{|F_\chi|^2}{M_P^2}X^\dag X\sim H^2X^\dag X
\ee
so a \emph{positive} K\"ahler potential coefficient results in a \emph{negative} coefficient for the soft mass term, potentially winning out over the previous contribution.  Since this soft mass is generated by the high scale SUSY breaking associated with inflation rather than the SUSY breaking associated with the MSSM, the result is independent of the scale appearing in the O'Raifeartaigh model itself.

Assuming the soft mass term {\em is} tachyonic, the effective potential is initially unstable around the origin and is only stabilised at
\be
X\sim\nbrack{HM^{n-3}}^{1/(n-2)}\quad\implies\quad
V_{\rm eff}(X)\sim\nbrack{H^{n-1}M^{n-3}}^{2/(n-2)}
\ee
with a choice of $n-2$ distinct minima corresponding to different choices of phase.  The condensate field quickly settles into one of these minima and remains there throughout inflation due to the large, Hubble induced damping term in eq.~\eqref{eq:CEeom} \cite{Dine:1995kz}.  Immediately after inflation the Hubble parameter evolves as $1/t$.  The minimum thus moves closer to the origin over time until $V_{\rm eff}(X)\sim\lambda^2f^2/16\pi^2$.  At this stage the low energy effective potential \eqref{eq:FDUeff} takes over\footnote{We shall assume there are no independent minima at large $X$, i.e.\ ones that do not require a negative mass term centred around the origin to be stable.  If such minima did exist, the condensate expectation value could remain near the cutoff scale and the theory would never flow into its low energy O'Raifeartaigh model description.  This effect is not of interest here, but could provide a novel mechanism for models with uplifted vacua \cite{Giveon:2009yu, Abel:2009ze, Koschade:2009qu, Kutasov:2009kb, Barnard:2009ir, Auzzi:2010wm, Maru:2010yx, Curtin:2010ku} to find themselves in a higher energy vacuum.}, the corresponding Hubble parameter and condensate expectation value being
\be\label{eq:CELambda}
H\sim\sbrack{\frac{1}{M^{n-3}}\nbrack{\frac{\lambda f}{4\pi}}^{n-2}}^{1/(n-1)}\quad\implies\quad
X\sim\Lambda\equiv\nbrack{\frac{\lambda f M^{n-3}}{4\pi}}^{1/(n-1)}.
\ee
Here, we have defined the parameter $\Lambda$ which will be important in all that follows.  This scale is parametrically between the cutoff scale of the O'Raifeartaigh model $M$ and the loop suppressed SUSY breaking scale $\sqrt{\lambda f/4\pi}$.  It actually turns out that the case $\Lambda^2<f$ is uninteresting so we will henceforth assume $f^2\ll\Lambda\ll M$.  It will also be convenient to recast the above value of the Hubble parameter in terms of $\Lambda$
\be
H\sim\frac{\lambda f}{4\pi\Lambda}.
\ee

\subsection{Rotation}

Below $H\sim \lambda f/4\pi\Lambda$ the condensate performs rotations about the minimum of the low energy effective potential \eqref{eq:FDUeff}.  Whether R-symmetry is spontaneously broken or not, the logarithmic regime of eq.~\eqref{eq:FDUeff} is independent of the condensate's phase.  An effective R-charge is thus conserved for the most part, resulting in approximately circular or elliptical motion.  The charge stored in the condensate is determined by the interplay between the various R-violating operators in the high energy effective potential \eqref{eq:CEVeff} when the rotation begins.  At the transition point between high and low energy regimes these are of comparable size to the R-preserving terms so can impart a sizeable `torque' on the condensate and bestow it with a large fractional charge\footnote{Relating R-charge to baryon number could thus lead to some interesting asymmetric dark matter scenarios such as those in refs.~\cite{Bell:2011tn, Cheung:2011if}.}.  The only subsequent source of R-symmetry violation is the small $X$ regime, hence we can think of the trajectory as being smooth and elliptical, but possibly getting a kick if it gets too close to the bottom of the potential well.

At first $X^2\sim\Lambda^2\gg f$ so the large $X$ limit of eq.~\eqref{eq:FDUeff} applies.  At any given time one could ignore the damping term in eq.~\eqref{eq:CEeom} and find a circular solution $X=X_ce^{i\nu t}$, for constant amplitude $X_c$ and frequency
\be\label{eq:CEomega}
\nu^2=\frac{1}{X_c}\pdiff{U_{\rm eff}}{X}=\frac{\lambda^2f^2}{16\pi^2X_c^2}.
\ee
Elliptical variants will be mentioned in section \ref{sec:RBform}.  Initially $X_c\sim\Lambda$ and $H\sim\lambda f/4\pi\Lambda$ so the damping coefficient is similar to the frequency and the motion is critically damped.  As the Hubble parameter continues to decrease, the damping follows suit and the motion becomes underdamped.  We thus expect the effect of the damping to be small, the condensate expectation value and frequency of rotation remaining around $\Lambda$ and $\lambda f/4\pi\Lambda$ respectively.

\subsection{Fragmentation\label{sec:CEfrag}}

Left to its own devices the damping would, eventually, force the condensate into the small $X$ regime of eq.~\eqref{eq:FDUeff} and it would continue to perform underdamped oscillations about $X=\expect{X}$ with frequency $m$ and magnitude $\sqrt{f}$.  However, over the domain $f\ll X^2<\Lambda^2$ the effective potential is logarithmic so increases slower than quadratically.  A condensate oscillating in this kind of potential behaves as matter with a negative pressure, i.e.\ it is unstable with respect to spatial perturbations \cite{Turner:1983he, Kusenko:1997si, Enqvist:1997si, Kasuya:1999wu, Kasuya:2000wx, Kasuya:2001hg, Multamaki:2002hv, Jokinen:2002xw}.

For circular rotations it is possible to see this explicitly and estimate the typical size of the fragments \cite{Kusenko:1997si, Kasuya:2001hg}.  Working in the underdamped regime discussed above one can write down the approximate solution
\be
X\approx\Lambda e^{i(\lambda f/4\pi\Lambda)t}.
\ee
Now consider fluctuations in the magnitude and phase of $X$, of the forms $\delta\xi=\delta\xi_0e^{\alpha t+ikx}$ and $\delta\theta=\delta\theta_0e^{\alpha t+ikx}$ respectively.  Unstable modes have $\alpha>0$.  Substituting into the equations of motion a non-trivial solution for $\delta\xi_0$ and $\delta\theta_0$ exists only if
\be
\alpha^4+2\nbrack{\frac{k^2}{a^2}+\frac{\lambda^2f^2}{16\pi^2\Lambda^2}}\alpha^2+\nbrack{\frac{k^2}{a^2}-\frac{\lambda^2f^2}{8\pi^2\Lambda^2}}k^2=0
\ee
which is satisfied for real, positive $\alpha$ and $k$ only if
\be
\nbrack{\frac{k^2}{a^2}-\frac{\lambda^2f^2}{8\pi^2\Lambda^2}}<0\quad\implies\quad
0<\frac{k}{a}<\frac{\lambda f}{2\pi\sqrt{2}\Lambda}.
\ee
The most amplified mode (largest $\alpha$) has a scale set by the upper limit of this inequality, so the fragments have typical size
\be\label{eq:CErfrag}
r\sim\frac{a}{k}\sim\frac{2\pi\sqrt{2}\Lambda}{\lambda f}.
\ee
Moving away from circular trajectories requires a numerical approach, but for a logarithmic effective potential it only changes the result by factors of order unity \cite{Kasuya:2001hg}.  Regardless, fragmentation occurs after the horizon size $H^{-1}$ surpasses the typical size of the fragments.  This has already happened when the condensate begins to rotate at $H\sim\lambda f/4\pi\Lambda$ so the process can take place immediately.

\subsection{Thermal effects\label{sec:CEtherm}}

So far we have ignored the fact that all of these processes take place in a thermal bath.  Below the reheat temperature $T_h$ the inflaton decays into light degrees of freedom.  This reheats the universe, at which point it can be considered a cooling plasma with temperature $T\sim\sqrt{HM_P}$.  Even beforehand the universe contains a dilute plasma with a lower temperature $T\sim(HM_PT_h^2)^{1/4}$ \cite{Kolb:1990vq}\footnote{Note that this temperature should {\em not} be associated with the temperature of the R-balls as they are {\em not} generally in thermal equilibrium.}.  The permitted values for the reheat temperature are constrained by gravitino cosmology.  If reheating occurs too early gravitino decay products disrupt nucleosynthesis or, if the gravitino is stable, its relic density may be too high \cite{Ellis:1982yb, Khlopov:1984pf, Ellis:1984eq, Kawasaki:1994af, Sarkar:1995dd}.  For an unstable gravitino (e.g.\ gravity mediation) one requires $T_h\lesssim10^6m_{3/2}$ whereas for a stable gravitino (e.g.\ gauge mediation) the bound is $T_h\lesssim10^{14}(m_{3/2}/\mbox{GeV})^{-2}$ GeV.  Additionally, there is a lower bound of 10 MeV placed on the reheat temperature to ensure that reheating happens before nucleosynthesis.

Any particle coupling to the condensate field at tree level acquires a mass of order $|\lambda X|$.  If this mass is greater than the plasma temperature, i.e.\ $|\lambda X|>T$, the state decouples from the condensate.  Otherwise the condensate is eroded.  The condensate expectation value goes like $\Lambda$ during fragmentation so one must have
\be\label{eq:CETfr}
\Lambda\gtrsim\frac{T_f}{\lambda}
\ee
for the process to proceed unhindered, where $T_f$ is the temperature of the universe at this time.  If the inequality is satisfied there remain thermally sensitive loop corrections to the the low energy effective potential \eqref{eq:FDUeff} to take into account \cite{Kasuya:2001hg}.  Couplings to heavy states yield the extra term\footnote{In addition one should include a temperature dependence in $\lambda$.  We shall assume this is small enough to omit here so as to avoid further complicating the discussion.}
\be
U_{\rm eff}(X,T)=\frac{\lambda^2T^4}{16\pi^2}\ln\nbrack{\frac{|X|^2}{T^2}}.
\ee
Unless $f>T^2$ this contribution dominates and the above expression should be used in place of eq.~\eqref{eq:FDUeff}.  However, since the form of the effective potential (which arises from loop corrections anyway) is not changed by the thermal contribution, its effects can be absorbed into the parameter $f$
\be\label{eq:CEfT}
f\longrightarrow f_T=\left\{\begin{array}{ll}
\widerow f & \mbox{for } T^2<f \\
\widerow T^2 & \mbox{for } T^2>f. \end{array}\right.
\ee
Other thermal effects exist but are expected to be small for the large R-balls we will be discussing \cite{Kusenko:1997si, Rummukainen:1998as} so will not be considered.

According to section \ref{sec:CEfrag} the condensate fragments at $H\sim\lambda f_T/4\pi\Lambda$ so, using the expressions given at the start of this section to relate the Hubble parameter to the plasma temperature, this equation breaks up into four possible domains depending on the relative sizes of the scales
\begin{align}
D_1:\quad T_f<T_h\,,\,T_f^2<f \quad\implies\quad & T_f^2/M_P\sim\lambda f/4\pi\Lambda \nonumber\\
D_2:\quad T_f>T_h\,,\,T_f^2<f \quad\implies\quad & T_f^4/M_PT_h^2\sim\lambda f/4\pi\Lambda \nonumber\\
D_3:\quad T_f>T_h\,,\,T_f^2>f \quad\implies\quad & T_f^4/M_PT_h^2\sim\lambda T_f^2/4\pi\Lambda \nonumber\\
D_4:\quad T_f<T_h\,,\,T_f^2>f \quad\implies\quad & T_f^2/M_P\sim\lambda T_f^2/4\pi\Lambda.
\end{align}
There are then three different solutions for the fragmentation temperature; the fourth solution is just the boundary between two other domains.  In each case the important parameters can be summarised as follows
\be\label{eq:CEtable}
\begin{array}{|l|ccc|}\hline
\widerow \mbox{Domain} & T_f & f_T & \Lambda_{\rm min} \\\hline\hline
\widerow D_1:\quad\Lambda>\max{\sbrack{\frac{\lambda fM_P}{4\pi T_h^2}\,,\,\frac{\lambda M_P}{4\pi}}} & \nbrack{\frac{\lambda fM_P}{4\pi\Lambda}}^{1/2} & f & \nbrack{\frac{fM_P}{4\pi\lambda}}^{1/3} \\
\widerow D_2:\quad\frac{\lambda fM_P}{4\pi T_h^2}>\Lambda>\frac{\lambda M_PT_h^2}{4\pi f} & \nbrack{\frac{\lambda fM_PT_h^2}{4\pi\Lambda}}^{1/4} & f & \nbrack{\frac{fM_PT_h^2}{4\pi\lambda^3}}^{1/5} \\
\widerow D_3:\quad\Lambda<\min{\sbrack{\frac{\lambda M_PT_h^2}{4\pi f}\,,\,\frac{\lambda M_P}{4\pi}}} & \nbrack{\frac{\lambda M_PT_h^2}{4\pi\Lambda}}^{1/2} & \frac{\lambda M_PT_h^2}{4\pi\Lambda} & \nbrack{\frac{M_PT_h^2}{4\pi\lambda}}^{1/3} \\\hline
\end{array}
\ee
where $\Lambda_{\rm min}$ is the value demanded by the inequality \eqref{eq:CETfr}.

\section{R-balls\label{sec:RB}}

Much as MSSM condensates end up as B-balls (or L-balls), our condensate collapses into extended, classical configurations with large charge, R-balls, which are non-topological solitons formed under the influence of the U(1) R-symmetry \cite{Coleman:1985ki, Kusenko:1997zq, Barnard:2010wk}.  Recall that the low energy effective potential \eqref{eq:FDUeff} preserves R-symmetry in the large $X$ regime whether or not it is spontaneously broken in the vacuum, hence R-balls form in {\em both} cases but have different decay properties depending on the eventual status of the symmetry.

R-ball solutions take the form $X(x,t)=X(x)e^{i\omega t}$, for a real parameter $\omega$ and a real function $X(x)$ that minimises
\be
\dimint{3}{x}\nbrack{\frac{1}{2}|\nabla X|^2+U_\omega(X)}
\ee
where a new potential
\be
U_\omega(X)\equiv U_{\rm eff}(X)-\frac{1}{2}\omega^2|X|^2
\ee
has been defined.  Assuming spherical symmetry the associated equations of motion are
\be\label{eq:RBeom}
\diff{^2X}{r^2}+\frac{2}{r}\diff{X}{r}-\diff{U_\omega}{X}=0
\ee
for a radial coordinate $r$.  One can always find a solution if $U_\omega(X)$ meets two criteria: it retains a stable minimum at $X=\expect{X}$ and there exists a non-zero $X$ such that $U_\omega(X)<U_\omega(\expect{X})$.  As long as $\omega<m$ the potential defined by eq.~\eqref{eq:FDUeff} satisfies both; $\omega^2X^2$ always beats the logarithm for large $X$.

\begin{figure}[!tb]
\begin{center}
\includegraphics[width=0.44\textwidth]{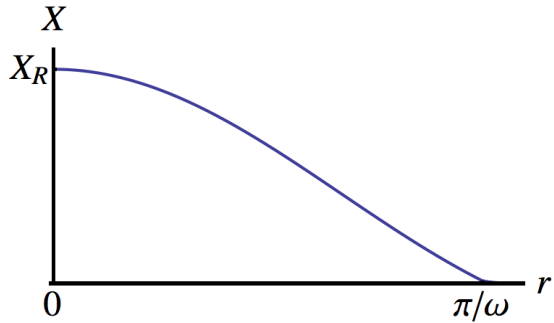}
\caption{A typical R-ball solution.  At small and large $r$ it has the forms $\sin{\omega r}/\omega r$ and $e^{-mr}$ respectively, with a width of about $\pi/\omega$.  The parameter $\omega$ scales like $R^{-1/4}$, the height like $R^{1/4}$ and the energy $R^{3/4}$.\label{fig:RBsol}}
\end{center}
\end{figure}

For the case at hand we can be more precise owing to the effective potential being so flat \cite{Dvali:1997qv, Copeland:2009as}.  Substituting eq.~\eqref{eq:FDUeff} into the equation of motion \eqref{eq:RBeom} yields two limits:
\be\label{eq:RBeom2}
X^{\prime\prime}+\frac{2}{r}X^\prime+\omega^2X=\left\{\begin{array}{ll}
\widerow m^2X & \mbox{for } X^2\ll f \\
\widerow \zeta\lambda^2f_T^2/16\pi^2X & \mbox{for } X^2\gg f.
\end{array}\right.
\ee
In the first, the equation is exactly soluble.  In the second, the right hand side can be ignored as $X$ is large, in which case the equation is again exactly soluble.  Choosing the solution that is finite at $r=0$, decays to zero as $r\rightarrow\infty$ and is continuous between the two regimes we find
\begin{align}
X & =\frac{X_R}{\omega r}\sin{\omega r}\quad\mbox{for } r<r_R\,, &
X & =\frac{X_R}{\omega r}\sin{\omega r_R}\,e^{(r_R-r)\sqrt{m^2-\omega^2}}\quad\mbox{for } r>r_R.
\end{align}
The R-ball's width is approximated by $r_R\approx\pi/\omega$, such that the cross over occurs near the first zero of $\sin{\omega r}$ where $X$ is arbitrarily small.  Note that a cross over to large $X$ is inevitable because the small $X$ solution diverges as $r\rightarrow0$.  When $\omega\ll m$ the R-ball is very large and the solution can be further simplified to find
\begin{align}\label{eq:RBsol}
X & =\frac{X_R}{\omega r}\sin{\omega r}\quad\mbox{for } r<r_R\,, &
X & =\frac{X_R}{\omega r_R}\sin{\omega r_R}\,e^{m(r_R-r)}\quad\mbox{for } r>r_R.
\end{align}
Evaluating the energy of this field configuration and choosing $\omega r_R\approx\pi$ gives
\be
E=\frac{\zeta\pi^2\lambda^2f_T^2}{12\omega^3}+\omega R+\cO(\pi-\omega r_R).
\ee
Minimising with respect to $\omega$ and explicitly evaluating the charge finally yields expressions for the R-ball parameters
\begin{align}\label{eq:RBparams}
\omega_R & \sim R^{-1/4}\sqrt{\lambda f_T} &
r_R & \sim\frac{R^{1/4}}{\sqrt{\lambda f_T}} &
X_R & \sim R^{1/4}\sqrt{\lambda f_T} &
E_R & \sim R^{3/4}\sqrt{\lambda f_T}
\end{align}
up to order one coefficients.  Putting it all together, a typical R-ball solution is illustrated in figure \ref{fig:RBsol}.

\subsection{Formation\label{sec:RBform}}

In section \ref{sec:CEfrag} it was shown that, for an initially circular trajectory, the condensate forms fragments of size $r\sim2\pi\sqrt{2}\Lambda/\lambda f_T$ shortly after it starts to rotate with magnitude $X_c\sim\Lambda$ and frequency $\nu_c\sim\lambda f_T/4\pi\Lambda$.  We can use these details to estimate the typical charge \cite{Kasuya:2001hg}.  The charge density during fragmentation is given by $2\nu_cX_c^2$ so the total charge per fragment is
\be
R\sim\frac{100\Lambda^4}{\lambda^2f_T^2}.
\ee
If each fragment collapsed into a single R-ball this would be their charge.  Actually, numerical simulations \cite{Kasuya:2001hg} suggest that R-balls form slightly after fragmentation in a logarithmic potential, leading to a reduced charge of
\be\label{eq:RBR}
R_c\sim\frac{\Lambda^4}{10\lambda^2f_T^2}.
\ee
Each fragment therefore contains about $10^3$ R-balls.

Using the energy \eqref{eq:RBparams} the energy density stored in R-balls just after formation, at $H\sim\lambda f_T/4\pi\Lambda$, can now be determined to be
\be\label{eq:RBOmega}
\rho_R(T_f)\sim\frac{\lambda^2f_T^2}{10}\quad\implies\quad
\Omega_R(T_f)=\frac{\rho_R(t_f)}{3H^2M_P^2}\sim\frac{10\Lambda^2}{M_P^2}
\ee
unless $\Lambda\sim M_P$ whereupon $\Omega_R\sim1$ and R-balls dominate the universe until they decay.   Of course, this result scales with the expansion of the universe.  If R-balls form above a temperature of $\sqrt{f}$ there is also lowering in density due to the decreasing value of $f_T$ \eqref{eq:CEfT}.  Treating R-balls as non-relativistic matter the density can be related to the scale factor via $\rho_R\sim a^{-3}$, implying that
\be\label{eq:RBOmegaex}
\Omega_R(T)\propto\left\{\begin{array}{ll}
\widerow \sqrt{f_T} & \quad\mbox{for matter domination} \\
\widerow \sqrt{f_T}/T & \quad\mbox{for radiation domination.} \end{array}\right.
\ee
It should be noted that there are no other processes to form R-balls once the condensate has fragmented so they cannot maintain thermal equilibirum.  The idea of solitosynthesis \cite{Griest:1989bq, Frieman:1989bx, Kusenko:1997hj} does not apply here as the goldstino carries R-charge and is lighter than the condensate field.

To generalise to more eccentric condensate trajectories one can define the parameter $\epsilon=\nu/\nu_c=4\pi\nu\Lambda/\lambda f_T$: the ratio of the angular velocity of the condensate to its maximal, circular value.  Equivalently it can be thought of as the fraction of the maximum possible charge that is stored in the condensate.  Clearly $\epsilon$ can lie anywhere in the range zero to one, zero corresponding to pure radial oscillations with vanishing net charge.  Naively one might expect the charge of the resultant R-balls to be given by $\epsilon R_c$.  This is mainly true.  However, once $\epsilon$ drops below about 0.06, both positive and negatively charged R-balls are formed in comparable quantities and the typical charge becomes constant \cite{Kasuya:1999wu, Kasuya:2000wx, Kasuya:2001hg}.

\subsection{Decay}

Bosonic decay modes do not exist when the vacuum of the low energy theory preserves R-symmetry (models which spontaneously break R-symmetry will be discussed shortly).  The R-ball is already the lowest energy scalar field configuration for a given charge and gauge bosons have R-charge zero.  However, several other decay modes {\em do} exist in most models; basically any light fermion with non-zero R-charge.  Even so decay is slower than one might think due to the Pauli exclusion principle.  Within an R-ball a Fermi pressure opposes the creation of fermions so, effectively, R-balls only evaporate from their surface.

Consider first the couplings in the superpotential \eqref{eq:FDW}.  In the absence of fine tuning one expects all particles to have tree level masses of order $\mu$, or zero if imposed by symmetry.  Since the condensate field mass $m$ is generated at one loop it is comparatively suppressed and only decays to the massless fermions are kinematically allowed.  Any massless fermion must be a null eigenvector of the fermionic mass matrix, given by $\mu$ in an R-symmetric model when $\expect{X}=0$.  Consequently its scalar partner is massless (see ref.~\cite{Komargodski:2009jf} or section \ref{sec:FD}) and the superfield must be a null eigenvector of $\lambda$ as well.  Therefore massless fermions are forbidden from coupling to the condensate field at tree level.  The condensate can only decay to the fermionic components of the $\varphi$'s at tree level if R-symmetry is spontaneously broken.

A second option for R-ball decay is the gravitino.  The goldstino has R-charge $+1$ and is originally exactly massless, but gets eaten by the gravitino to pick up a mass $m_{3/2}\sim f_T/M_P$.  Decay then proceeds via goldstino interaction terms \cite{Fayet:1977vd, Fayet:1979yb}
\be\label{eq:SFg}
\cL\supset\frac{m^2}{f_T}X^\dag\tilde{G}\tilde{G}
\ee
with $\tilde{G}$ denoting the gravitino.  As the condensate field mass is generated at one loop the coupling is effectively loop suppressed.  Loop decays to other light fermions (that may be external to the hidden sector) should therefore be considered too.  One may also expect R-violating modes induced either by gravity, which is not excepted to respect any global symmetries, or the microscopic theory, which may break R-symmetry.  These decay channels are suppressed by at least the cutoff scale so are not expected to be important.

Rigorously estimating the lifetime of an R-ball in the thick wall limit is a difficult task \cite{Multamaki:1999an} so here we shall take a simpler approach.  The thin wall limit \cite{Cohen:1986ct} is not appropriate for the R-ball solution given in \eqref{eq:RBsol} but the principles applied to bosonic decay modes in ref.~\cite{Enqvist:1998en} are.  First consider tree level decays such as those to gravitinos.  For a fermion coupling directly to the condensate field with strength $g$ (equal to $m^2/f_T$ for gravitinos) the penetration width is $1/gX$.  The fermion is produced inside the R-ball at some radius $r$ so one must have
\be
\frac{1}{gX}>r_R-r
\ee
if it is to escape.  Otherwise the Fermi pressure prevents decay.  Since we are dealing with the interior of the R-ball, the large $X$ component of eq.~\eqref{eq:RBsol} is appropriate and the inequality becomes
\be
\frac{\pi-\omega_Rr}{\omega_Rr}\sin{\omega_Rr}<\frac{\omega_R}{gX_R}\sim\frac{1}{g\sqrt{R}}
\ee
using $\omega_Rr_R\approx\pi$ and, for the last term, eq.~\eqref{eq:RBparams}.  When the charge is large (specifically $R\gg g^{-2}$) this inequality can only satisfied around the surface of the R-ball, $r=r_R$.  One thus finds
\be\label{eq:RBdotRf}
\dot{R}\approx-4\pi\omega_R\dimintlim{}{r}{r_R}{\infty}r^2 \Gamma X^2
\ee
where $\Gamma$ is the decay rate at a given point inside the R-ball and $\omega_RX^2$ is the local charge density.

Decreasing the charge of the R-ball by two (a single condensate quanta) liberates energy of order $\omega_R$, yet fermions coupling to the condensate field at tree level gain local masses of order $gX$.  Unless $gX<\omega_R$ a local, tree level decay is kinematically forbidden and decay can only occur via heavy particle loops.  On the surface $X=\sqrt{f_T}$ by definition so the inequality is already satisfied there as long as $g^2<\lambda R^{-1/2}$.  If not, the crossover occurs at a radius
\be\label{eq:RBgamma}
\gamma r_R\sep{for}\gamma\sim1+\frac{1}{4mr_R}\ln{\nbrack{\frac{Rg^4}{\lambda^2}}}
\ee
where the expressions for the small $X$ component of eq.~\eqref{eq:RBsol} and the R-ball parameters \eqref{eq:RBparams} have been utilised.  Large R-balls have $r_R\approx\pi/\omega_R\gg1/m$ so the second term is subdominant and we can approximate $\gamma$ by one, meaning that $gX<\omega_R$ everywhere from the surface outwards.  The decay rate is therefore constant at its tree level value $\Gamma\approx g^2\omega_R/4\pi$.  Essentially the R-ball solution decays very rapidly beyond the surface, so if $gX<\omega_R$ is not already satisfied there it soon will be.  Putting it all together we find
\be
\dot{R}\sim-\frac{g^2f_T\omega_R^2}{4m^3}\nbrack{1+2mr_R+2m^2r_R^2}\sim-\frac{\pi^2g^2f_T}{2m}
\ee
where the last equality again follows from the fact that $mr_R\approx\pi m/\omega_R\gg1$ for large charge\footnote{In more detail, the decay rate starts off constant but, as the charge of the R-ball decreases, the other two terms grow and decay speeds up.  Consequently R-balls decay more quickly with decreasing size, as was seen numerically for Q-balls in ref.~\cite{Multamaki:1999an}.}.

Loop decays tell a similar story.  As long as the mass of the heavy particle propagating around the loop is greater than the mass it acquires from its tree level coupling to the condensate, the result $\Gamma\approx g^2\omega_R/4\pi$ is unchanged.  One simply replaces $g$ with the appropriate effective coupling.  Fortunately this will always be the case just beyond the surface for the reasons mentioned above.  Coupling fields in the loop to light fermions with strength $h$ thus gives a decay rate $\Gamma\approx g^2h^4\omega_R^3/16\pi^2\mu^2$ and, subsequently
\be
\dot{R}\sim-\frac{\pi\lambda g^2h^4f_T^2}{8m\mu^2}R^{-1/2}.
\ee

When $T_f^2<f$ the tree level decay rate is easily integrated to find an approximate lifetime
\be\label{eq:RBtauRSt}
\tau_{\rm RS}\sim\frac{2m}{\pi^2g^2f}R\sim\frac{\Lambda^4}{10^3g^2f^2\mu}
\ee
using eq.~\eqref{eq:RBR} to set the initial charge and eq.~\eqref{eq:FDm} to eliminate $m$.  If loop decays dominate this result becomes
\be\label{eq:RBtauRSl}
\tau_{\rm RS}\sim\frac{16m\mu^2}{3\pi\lambda^3h^4f_T^2}R^{3/2}\sim\frac{\Lambda^6\mu}{10^3\lambda^2g^2h^4f^4}.
\ee
Otherwise the lifetime depends on the thermal history of the universe.  Decay is quicker and the initial charge is smaller when the temperature is greater than $\sqrt{f}$, due to eq.~\eqref{eq:CEfT}.  A lower bound is thus found by fixing $f_T$ at $T_f^2$, whereupon one can simply replace $f$ with $T_f^2$ in the above expressions.

\subsection{Low energy theories without R-symmetry\label{sec:RBdRB}}

In O'Raifeartaigh models that spontaneously break their R-symmetry the initial situation remains broadly similar.  Immediately after fragmentation the condensate moves in the logarithmic regime of the effective potential \eqref{eq:FDUeff}.  This conserves R-charge so supports the formation of R-balls as before.  The main difference is that there may now be R-violating bosonic decay modes available, which are not stifled by a Fermi pressure so can shorten the lifetime.  Expressions for the fermionic decay rate are unaffected (we assumed nothing about charge conservation in the previous section short of establishing what decay modes were allowed) and the effect of bosonic modes can be estimated in a similar manner \cite{Enqvist:1998en}.

The key difference is that the lower limit of the integral \eqref{eq:RBdotRf} is zero in the absence of Fermi pressure, so tree level decays acquire an extra contribution from the interior of the R-ball
\be
\Delta\dot{R}\approx-4\pi\omega_R\dimintlim{}{r}{0}{r_R}r^2\Gamma X^2\approx-g^2\omega_R^2\nbrack{\frac{g^2\omega_R^2}{4\pi}\dimintlim{}{r}{0}{\gamma r_R}r^2+\dimintlim{}{r}{\gamma r_R}{r_R}r^2X^2}.
\ee
Between $r=0$ and $\gamma r_R$ decay is at one loop via particles of mass $gX$, giving $\Gamma\approx g^4\omega_R^3/16\pi^2X^2$, whereas between $\gamma r_R$ and the surface we instead have $\Gamma\approx g^2\omega_R/4\pi$.  Due to the arguments associated with eq.~\eqref{eq:RBgamma} one expects $\gamma\le1$, with equality when $g^2>\lambda R^{-1/2}$.  For $g^2<\lambda R^{-1/2}$ the crossover occurs inside the R-ball where the solution changes more slowly so one expects the deviation from $\gamma\approx1$ to be significant.  Substituting in the large $X$ component of eq.~\eqref{eq:RBsol} and $r_R\approx\pi/\omega_R$ the increase is given by
\ba
\Delta\dot{R} &\sim& -\frac{\pi^2\gamma^3g^4\omega_R}{12}-\frac{g^2X_R^2}{4\omega_R}(2\pi(1-\gamma)+\sin{2\pi\gamma}) \nonumber\\
&\sim& -\frac{g^2\sqrt{\lambda f_T}}{4}R^{3/4}(2\pi(1-\gamma)+\sin{2\pi\gamma}).
\ea
Large $R$ means small $\omega_R$ (specifically $\omega_R\ll m$, implying that $\omega_R\ll f_T/m$) whereupon the first term on the first line can be neglected.  The second term, however, is extremely significant for large R-balls as it dominates the overall decay rate.

Loop decays to bosons are similarly enhanced, contributing
\be
\Delta\dot{R}\sim-\frac{g^2h^4(\lambda f_T)^{3/2}}{16\pi\mu^2}R^{1/4}(2\pi(1-\gamma)+\sin{2\pi\gamma}).
\ee
The positive exponent attached to $R$ means that even one loop bosonic channels can be more important than tree level fermionic ones if the charge is large enough.  However, note that if $g^2>\lambda R^{-1/2}$ (or $g>\mu/\sqrt{f_T}$ for loop decays) $\gamma\approx1$ as before.  Decay to bosons is then suppressed everywhere inside the R-ball due to the mass they acquire from condensate field couplings.  Evaporation is only from the surface as per the R-symmetric case and there is no longer a significant enhancement.

Assuming that $g^2<\lambda R^{-1/2}$ and $T_f^2<f$ the revised lifetime for tree level decay modes is easily evaluated to be
\be\label{eq:RBtauRBt}
\tau_{\rm RB}\sim\frac{4}{g^2\sqrt{\lambda f}}R^{1/4}\sim\frac{\Lambda}{\lambda g^2f}
\ee
using eq.~\eqref{eq:RBR} to set the initial charge and assuming that $(2\pi(1-\gamma)+\sin{2\pi\gamma})/4\sim1$.  For loop decays one finds
\be\label{eq:RBtauRBl}
\tau_{\rm RB}\sim\frac{64\pi\mu^2}{3\lambda^2h^4(\lambda f)^{3/2}}R^{3/4}\sim\frac{10\Lambda^3\mu^2}{\lambda^3g^2h^4f^3}.
\ee
Just as in the R-symmetric case one can find a lower bound on the lifetime when the formation temperature is greater than $\sqrt{f}$ by replacing $f$ with $T_f^2$ in these expressions.  Of course, all of this assumes the existence of light bosons that the condensate is able to decay.  It could be that all bosons are too heavy.  Or the model parameters could conspire to enable the fermionic decay rate to overtake the bosonic one.  In either case the lifetime reverts to that given in the previous section.

\section{R-ball phenomenology\label{sec:RP}}

The cosmological behaviour of R-balls is wide ranging.  To give a general overview of the features various models can exhibit, this section will focus mainly on O'Raifeartaigh models obeying the following criteria.  The strength of all couplings in the superpotential \eqref{eq:FDW} will be taken to be of order $\lambda$, whereas `naturalness' suggests that the two scales in the superpotential \eqref{eq:FDW} should be similar, i.e.\ $\mu^2\sim f_T$, accounting for thermal corrections with eq.~\eqref{eq:CEfT}\footnote{Increasing $\mu$ above this scale increases the tree level decay rate but decreases the one loop decay rate.}.  We will further assume that all degrees of freedom in the hidden sector are heavy with masses $\mu$, other than the condensate field.

R-ball formation takes place in O'Raifeartaigh models whenever the condensate field (the scalar partner of the Goldstino) picks up a tachyonic soft mass due to couplings to the inflaton.  When it does the characteristic R-ball scale $\Lambda$, defined in eq.~\eqref{eq:CELambda}, is parametrically between the cutoff scale of the model and the SUSY breaking scale.  The temperature of the universe at formation \eqref{eq:CEtable} and typical charge \eqref{eq:RBR} of large R-balls are then given in the following table
\be
\begin{array}{|c||ccc|}\hline
\widerow \mbox{Domain} & D_1 & D_2 & D_3 \\\hline
\widerow T_f & \nbrack{\frac{\lambda fM_P}{4\pi\Lambda}}^{1/2} & \nbrack{\frac{\lambda fM_PT_h^2}{4\pi\Lambda}}^{1/4} & \nbrack{\frac{\lambda M_PT_h^2}{4\pi\Lambda}}^{1/2} \\
\widerow R & \frac{\Lambda^4}{10^5\lambda^2f^2} & \frac{\Lambda^4}{10^5\lambda^2f^2} & \frac{\Lambda^6}{10^3\lambda^4M_P^2T_h^4} \\\hline
\end{array}.
\ee
Domain boundaries are determined by when R-ball formation takes place relative to the decay of the inflation, and whether thermal effects dominate the effective potential during this process.  They are defined by
\begin{align}\label{eq:RPdoms}
D_1:\quad T_f<T_h\,,\,T_f^2<f \quad\implies\quad & \Lambda>\max{\sbrack{\frac{\lambda fM_P}{4\pi T_h^2}\,,\,\frac{\lambda M_P}{4\pi}\,,\,\nbrack{\frac{fM_P}{4\pi\lambda}}^{1/3}}} \nonumber\\
D_2:\quad T_f>T_h\,,\,T_f^2<f \quad\implies\quad & \frac{\lambda fM_P}{4\pi T_h^2}>\Lambda>\max{\sbrack{\frac{\lambda M_PT_h^2}{4\pi f}\,,\,\nbrack{\frac{fM_PT_h^2}{4\pi\lambda^3}}^{1/5}}} \nonumber\\
D_3:\quad T_f>T_h\,,\,T_f^2>f \quad\implies\quad & \min{\sbrack{\frac{\lambda M_PT_h^2}{4\pi f}\,,\,\frac{\lambda M_P}{4\pi}}}>\Lambda>\nbrack{\frac{M_PT_h^2}{4\pi\lambda}}^{1/3}.
\end{align}
where $T_h$ denotes the reheat temperature of the universe after inflaton decay.  Outside of these domains the condensate is eroded before R-balls are able to form.  For gravity mediated SUSY breaking the reheat temperature must lie in the range $10\mbox{ MeV}\lesssim T_h\lesssim10^6m_{3/2}$ whereas for gauge mediation the bound is $10\mbox{ MeV}\lesssim T_h\lesssim10^{14}(m_{3/2}/\mbox{GeV})^{-2}\mbox{ GeV}$.

The initial energy density stored in R-balls \eqref{eq:RBOmega} is given by
\begin{align}\label{eq:RPOmegaf}
\Omega_R(T_f)\sim\frac{10\Lambda^2}{M_P^2}
\end{align}
(or one if $\Lambda\sim M_P$) and scales with the subsequent expansion of the universe \eqref{eq:RBOmegaex} as
\be\label{eq:RPOmega}
\Omega_R(T)\sim\left\{\begin{array}{ll}
\widerow 10\Lambda^2/M_P^2 & \quad\mbox{for } T_f>T>T_h \\
\widerow (T_h/T)(10\Lambda^2/M_P^2) & \quad\mbox{for } T_h>T>T_e \\
\widerow (T_h/T_e)(10\Lambda^2/M_P^2) & \quad\mbox{for } T_e>T \end{array}\right.
\ee
where $T_e\sim1\mbox{ eV}$ is the usual temperature at which the universe becomes matter dominated.  If $T_f<T_h$ one replaces $T_h$ with $T_f$ in the second equation and, for $\Lambda\in D_3$, the density is multiplied by $\sqrt{f_T}/T_f$ due to the temperature dependence \eqref{eq:CEfT} of the R-ball energy \eqref{eq:RBparams}.

From here on the analysis becomes strongly dependent on the particulars of the model and the available decay modes.  These in turn are specific to the mechanism chosen to mediate SUSY breaking.  The main examples we will consider are gravity and gauge mediated SUSY breaking, for preserved and spontaneously broken R-symmetry, both for strongly and weakly coupled superpotentials.  Some general features that crop in these models are as follows.

Perhaps of most immediate interest is the idea that R-balls are long lived and still exist today, contributing to the dark matter density of the universe.  According to eqs.~\eqref{eq:RBtauRSt}, \eqref{eq:RBtauRSl}, \eqref{eq:RBtauRBt} and \eqref{eq:RBtauRBl} long lifetimes correspond to a high cutoff scale (or an exact R-symmetry), a small SUSY breaking scale (or equivalently a small gravitino mass) and/or a weakly coupled condensate field.  If sufficiently long lived, non-relativistic R-balls behave as cold dark matter with a density given by eq.~\eqref{eq:RPOmega}.  One must find $\Omega_R(T_e)\le0.22$ if they are not to exceed the total observed value.  Note that light gravitinos, one of the requirements for R-ball dark matter, are specific to gauge mediated SUSY breaking, hence one does not expect this kind of dark matter in models of gravity mediation.

If not sufficiently long lived to survive until the present day R-balls must, obviously, decay.  The effect this has on the visible sector varies and could in principle result in either heating or cooling of the universe.  Each charge two quanta of condensate that decays carries away energy
\be\label{eq:RPDeltaE}
\Delta E\sim\left\{\begin{array}{ll}
\widerow \lambda f/\Lambda & \quad\mbox{for } \Lambda\in D_1,D_2 \\
\widerow \lambda f_T^{1/2}T_f/\Lambda & \quad\mbox{for } \Lambda\in D_3 \end{array}\right.
\ee
using eqs.~\eqref{eq:RBparams} and \eqref{eq:RBR}.  Regardless of what the decay products actually are, this sets their maximum mass and characteristic temperature.

Since R-balls are not in thermal equilibrium with the rest of the universe their decay products will initially be out of equilibrium as well.  Large R-ball density \eqref{eq:RPOmega} during decay thus results in a secondary reheating, or cooling of the universe.  An immediate consequence is that, if one occurs, an R-ball dominated epoch decouples the generation of visible sector matter and radiation from the dynamics of the inflaton.  R-ball, rather than inflaton, decay can be responsible for the present contents of the universe, with the new reheat temperature $\Delta E$ obeying the constraints applied to the original one $T_h$.  If, on the other hand, the R-ball density is small the temperature of the universe is unchanged.  The decay products are brought into equilibrium with everything else, unless they are incapable of maintaining thermal equilibrium in which case they simply update the relevant relic abundance by a small amount.  Either way the scenario is of limited interest.

R-symmetric models permit decays to light fermions alone.  The only such fermion that can couple to the condensate field at tree level is the gravitino but, owing to the loop suppressed mass of the condensate field, decays at one loop may also be important.  This is particularly so in the absence of light hidden sector states whereupon these decays are to visible sector gauginos (which have R-charge $+1$).  When the O'Raifeartaigh model spontaneously breaks R-symmetry R-balls can also decay at one loop to visible sector gauge bosons, sfermions or even fermions.  Alternatively, if light hidden sector degrees of freedom are included, one generally finds an increased decay rate as R-balls can decay directly to said fields rather than via loops or gravitinos.  However, the observable effects in this scenario are highly model specific and depend on the exact properties of these new degrees of freedom so will not be discussed here.

\subsection{Gravity mediation}

In pure gravity mediation the O'Raifeartaigh model communicates with the visible sector only through gravitational interactions.  Gravitinos couple to the condensate at tree level with strength $g\sim m^2/f_T\sim\lambda^4/16\pi^2$ so lead to an R-ball lifetime \eqref{eq:RBtauRSt}
\be\label{eq:RPtaugrRS}
\tau_{\rm RS}\sim\left\{\begin{array}{ll}
\widerow 10\Lambda^4/\lambda^8(m_{3/2}M_P)^{5/2} & \quad\mbox{for } \Lambda\in D_1,D_2 \\
\widerow 10\Lambda^4/\lambda^8T_f^5 & \quad\mbox{for } \Lambda\in D_3 \end{array}\right.
\ee
in R-symmetric vacua, where this is the only available decay mode.  The only light states accessible to the condensate at one loop are those in the visible sector and it must be gravitinos propagating around the loop.  Loop decay rates are therefore negligible as they contain higher orders of the already small gravitino couplings.

Spontaneously broken R-symmetry opens up the possibility of decay to visible sector gauge bosons via gravitino loops.  This decay rate is only significant if the R-ball interior contributes, i.e.\ $g<1$ (see section \ref{sec:RBdRB}), which is always true for the small gravitino coupling.  Gravitino couplings to gauge bosons go like $h\sim m_{3/2}/M_P$ leading to the shortened R-ball lifetime \eqref{eq:RBtauRBl}
\be
\tau_{\rm RB}\sim\left\{\begin{array}{ll}
\widerow 10^5\Lambda^3M_P^2/\lambda^{11}m_{3/2}^6 & \quad\mbox{for } \Lambda\in D_1,D_2 \\
\widerow 10^5\Lambda^3T_f^4/\lambda^{11}m_{3/2}^8 & \quad\mbox{for } \Lambda\in D_3 \end{array}\right.
\ee
provided this expression evaluates to less than \eqref{eq:RPtaugaRS}, otherwise decay remains to gravitinos.

\begin{figure}[!tb]
\begin{center}
\includegraphics[width=0.66\textwidth]{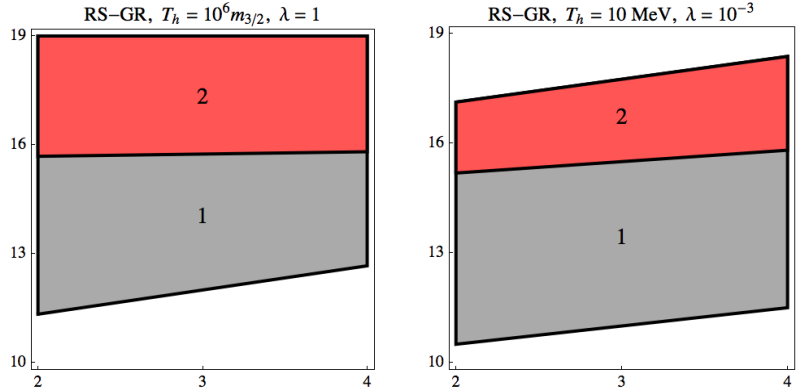}
\caption{R-balls in R-symmetric, gravity mediated SUSY breaking as a function of $\log{(m_{3/2}/\mbox{GeV})}$ (horizontal axis) and $\log{(\Lambda/\mbox{GeV})}$ (vertical axis) for various values of $\lambda$ and $T_h$.  In region 1 they have no significant effects.  In region 2 they reheat the universe by decaying to gravitinos.  Outside the shaded region R-balls do not form due to thermal effects, or they contribute more than the observed cold dark matter density.\label{fig:RPgr}}
\end{center}
\end{figure}

We can use these lifetimes to slice up the O'Raifeartaigh model parameter space ($m_{3/2}$, $\Lambda$, $\lambda$) into different regions of interest.  It actually turns out that R-balls in gravity mediated SUSY breaking do not have a wide range of phenomenological consequences; they can only result in secondary reheating of the universe by gravitinos.  This is because loop couplings remain sufficiently suppressed over the entire parameter space to favour gravitino decays, and the high SUSY breaking scale tends to keep their lifetime short.  Weak coupling does lead to a small region supporting long lived R-balls, but the density there is too high to be compatible with the observed cold dark matter density.  Some R-symmetric examples are given in figure \ref{fig:RPgr}.  R-breaking O'Raifeartaigh models still result in R-ball formation, but their density is typically too low to have any interesting consequences.

\subsection{Gauge mediation}

A combination of messenger loops and a low gravitino mass in gauge mediation lead to a much more interesting phenomenology.  Although gravitinos are still the only light fermion coupling to the condensate field at tree level, there are now one loop couplings via messengers to visible sector gauge fields.  These loops are much more significant than gravitino loops as messengers couple to the condensate field with strength $g\sim\lambda$, then through visible sector gauge couplings with strength $h\sim1$.  If gauginos are kinematically accessible, i.e. $\Delta E\gtrsim1\mbox{ TeV}$ \eqref{eq:RPDeltaE}, and decay to them is faster than to gravitinos, the R-ball lifetime \eqref{eq:RBtauRSl} is
\be\label{eq:RPtaugaRS}
\tau_{\rm RS}\sim\left\{\begin{array}{ll}
\widerow \Lambda^6/10^3\lambda^4(m_{3/2}M_P)^{7/2} & \quad\mbox{for } \Lambda\in D_1,D_2 \\
\widerow \Lambda^6/10^3\lambda^4T_f^7 & \quad\mbox{for } \Lambda\in D_3 \end{array}\right.
\ee
for an R-symmetric O'Raifeartaigh model.  Otherwise the lifetime is determined by the decay rate to gravitinos as in eq.~\eqref{eq:RPtaugrRS}.

R-breaking O'Raifeartaigh models are even more sensitive to loop decays.  Now visible sector gauge bosons are always accessible and dominate decay if either eq.~\eqref{eq:RPtaugaRS} or
\be
\tau_{\rm RB}\sim\left\{\begin{array}{ll}
\widerow 10\Lambda^3/\lambda^5m_{3/2}^2M_P^2 & \quad\mbox{for } \Lambda\in D_1,D_2 \\
\widerow 10\Lambda^3/\lambda^5T_f^4 & \quad\mbox{for } \Lambda\in D_3. \end{array}\right.
\ee
(using eq.~\eqref{eq:RBtauRBl}) is less than eq.~\eqref{eq:RPtaugrRS}.  In either case R-balls preferentially evaporate to gauge bosons rather than fermions due to the extra contribution from the interior.

\begin{figure}[!tb]
\begin{center}
\includegraphics[width=0.99\textwidth]{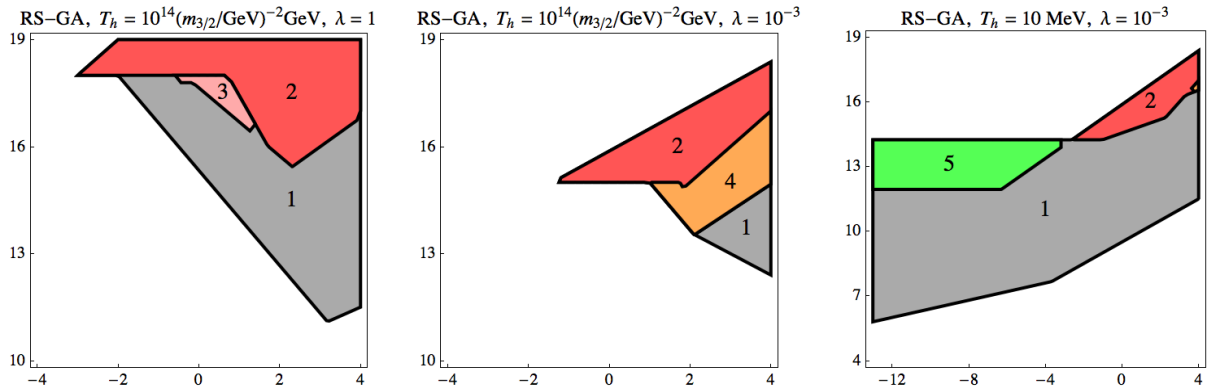}
\caption{R-balls in R-symmetric, gauge mediated SUSY breaking as a function of $\log{(m_{3/2}/\mbox{GeV})}$ (horizontal axis) and $\log{(\Lambda/\mbox{GeV})}$ (vertical axis) for various values of $\lambda$ and $T_h$.  In region 1 they have no significant effects.  In region 2/3 they reheat/cool the universe by decaying to gravitinos.  Region 4 corresponds to reheating by gauginos and region 5 to R-ball dark matter.  Outside the shaded region R-balls do not form due to thermal effects, contribute more than the observed cold dark matter density or their decay violates the bounds on the reheat temperature.\label{fig:RPgaRS}}
\end{center}
\end{figure}

The parameter space ($m_{3/2}$, $\Lambda$, $\lambda$) can be split as for gravity mediation but we now find more variation in the results.  R-symmetric O'Raifeartaigh models demonstrate both gravitino reheating and cooling, gaugino reheating and R-ball dark matter.  Dark matter corresponds to a small original reheat temperature so as to minimise the boost given to the density \eqref{eq:RPOmega}.  Otherwise the regions of parameter space coincident with long lived R-balls tend to be associated with too high a density.  Furthermore the gravitino mass is small in these regions so as to maximise the lifetime.

For most of the range in which R-balls live long enough the gravitino mass is less than a few keV.  Its contribution to the overall dark matter density is therefore small \cite{Pagels:1981ke}, but R-balls can easily account for the entire $\Omega_{\rm DM}\approx0.22$.  If they do (the top of region 5 in figure \ref{fig:RPgaRS}) their charge, size and energy \eqref{eq:RBparams} are in the ranges
\begin{align}\label{eq:RPdm}
10^{29}\lesssim R\lesssim10^{49}, &&
10^{-15}\mbox{ m}\lesssim r_R\lesssim10^{-4}\mbox{ m,} &&
10^{28}\mbox{ GeV}\lesssim E_R\lesssim10^{38}\mbox{ GeV.}
\end{align}
Individual R-balls can therefore be anything from fermi to micro scale and, regardless of size, are very dense objects.  Outside of the dark matter region gravitino decay is the norm for R-symmetric models due to the kinematic constraints on forming gauginos, but there are regions of parameter space favouring evaporation to gauginos.  Three examples of R-balls in R-symmetric, gauge mediated SUSY breaking are provided in figure \ref{fig:RPgaRS}.

\begin{figure}[!tb]
\begin{center}
\includegraphics[width=0.66\textwidth]{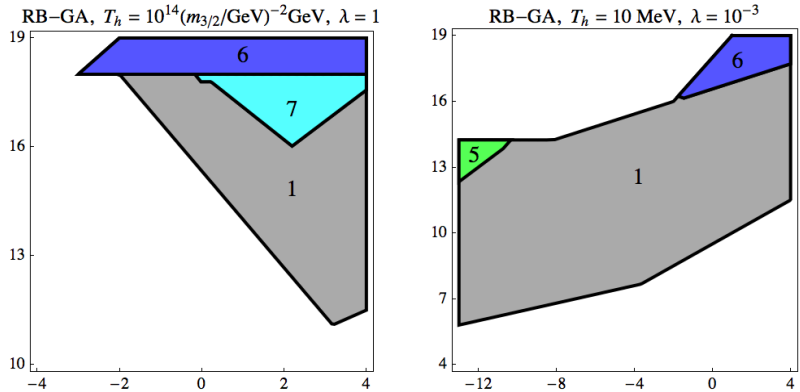}
\caption{R-balls in R-breaking, gauge mediated SUSY breaking as a function of $\log{(m_{3/2}/\mbox{GeV})}$ (horizontal axis) and $\log{(\Lambda/\mbox{GeV})}$ (vertical axis) for various values of $\lambda$ and $T_h$.  In region 1 they have no significant effects.  Region 5 corresponds to R-ball dark matter.  In region 6/7 they reheat/cool the universe by decaying to visible sector gauge bosons.  Outside the shaded region R-balls do not form due to thermal effects, contribute more than the observed cold dark matter density or their decay violates the bounds on the reheat temperature.\label{fig:RPgaRB}}
\end{center}
\end{figure}

O'Raifeartaigh models with spontaneously broken R-symmetry have a somewhat different phenomenology.  In all regions where R-ball decay has a significant impact on the evolution of the universe decay is to visible sector gauge bosons, and can result in either heating or cooling.  Cooling takes place if the original reheat temperature is high and $\Lambda\lesssim10^{18}\mbox{ GeV}$ (the decreased lifetime wins out against the increased energy), whereas heating occurs elsewhere.  There also remains a small region supporting R-ball dark matter, with parameters skewed towards the large R-balls of eq.~\eqref{eq:RPdm}.  Some example of R-balls in R-breaking, gauge mediated SUSY breaking are shown in figure \ref{fig:RPgaRB}.

\subsection{Detecting R-balls}

Experimentally, dark matter R-balls would be challenging to observe.  They cannot be produced in colliders, but one might hope for a dominant decay mode to visible sector particles that can be observed in some other type of experiment.  Gravitino decays and any decays to the hidden sector are thus ruled out, but decays at one loop via the messengers of gauge mediation are a possible candidate.  If R-symmetry is preserved the energy released in the decay of a single condensate quanta \eqref{eq:RPDeltaE} must be sufficient to produce a pair of gauginos, but if R-symmetry is spontaneously broken R-balls can always decay to massless visible sector gauge bosons.  From eq.~\eqref{eq:RPdm} the expected energy range for R-ball dark matter is
\be
10^{-11}\mbox{ GeV}\lesssim \Delta E\lesssim10^{-1}\mbox{ GeV}
\ee
which is well below the gaugino mass.  Ergo dark matter R-balls can only be observed through decay to visible sector particles if R-symmetry is spontaneously broken.

Q-ball detection techniques, as studied in refs.~\cite{Dvali:1997qv, Kusenko:1997it, Kusenko:1997vp, Kusenko:2004yw, Kusenko:2005du, Arafune:2000yv, Takenaga:2006nr, Kusenko:2009iz}, do not apply to R-balls as the condensate field is not charged under the visible sector gauge group.  One could perhaps search for them using direct detection experiments though.  Visible sector matter will scatter elastically off R-balls at one loop, through penguin diagrams containing messengers, for example.  Owing to the classical nature of R-balls this process is likely to be somewhat non-standard and could produce a distinctive signature.  The details are left for future work.

Evidence for R-balls decaying before the present day would surely necessitate the inception of a test with a more cosmological nature.  Their formation and decay are potentially significant events in the evolution of the universe so may well have left an imprint on some large scale, cosmological observable.  However, the details of such a test are beyond the scope of this work.

\section{Summary}

Condensates forming along flat directions of O'Raifeartaigh models can have a significant impact on the evolution of the universe.  They are somewhat generic, emerging in any model where the flat direction acquires a tachyonic soft mass through couplings to the inflaton.  When a condensate does form it eventually fragments into non-topological solitons with conserved R-charge, known as R-balls.  These objects are large, classical configurations and allow an approximate, analytical description.  Formation is insensitive to whether or not R-symmetry is spontaneously broken, but decay is not.

Depending on the scale up to which the O'Raifeartaigh model is valid, the scale of SUSY breaking and the strength of tree level couplings in the superpotential R-balls result in a variety of phenomena (figures \ref{fig:RPgr}, \ref{fig:RPgaRS} and \ref{fig:RPgaRB}).  In gravity mediated SUSY breaking they can reheat the universe through gravitino decays.  In gauge mediation they provide a good dark matter candidate if stable, or decay to gravitinos, gauginos or visible sector gauge bosons, either reheating or cooling the universe.  Both mediation mechanisms enable one to decouple the generation of visible sector matter from inflaton dynamics, instead using R-balls to reheat the universe.

\phantom{woooo! spooky!}

\noindent\textbf{Acknowledgements:} I would like to thank Steven Abel and Mark Goodsell for helpful comments and suggestions.  This work was supported by an STFC Postgraduate Studentship.

\bibliographystyle{JHEP-2}
\bibliography{../masterbib}
\end{document}